\documentstyle[prd,aps,12pt]{revtex}

\def\beq{\begin{equation}}
\def\eeq{\end{equation}}
\def\T{{\cal T}}
\def\r{{\bf r}}
\def\lsim{\
  \lower-1.2pt\vbox{\hbox{\rlap{$<$}\lower5pt\vbox{\hbox{$\sim$}}}}\ }
\def\gsim{\
  \lower-1.2pt\vbox{\hbox{\rlap{$>$}\lower5pt\vbox{\hbox{$\sim$}}}}\ }

\begin{document}
\tightenlines

\title{Newtonian limit of conformal gravity}

\author{ O.~V.~Barabash$^a$ and
Yu.~V.~Shtanov$^b$\footnote{shtanov@ap3.bitp.kiev.ua}}

\address{$^a$Department of Physics, Shevchenko National University, Kiev
252022, Ukraine \\ 
$^b$Bogolyubov Institute for Theoretical Physics, Kiev 252143, Ukraine\\}

\date{June 20, 1999}

\maketitle

\bigskip

\begin{abstract}
We study the weak-field limit of the static spherically symmetric solution of
the locally conformally invariant theory advocated in the recent past by
Mannheim and Kazanas as an alternative to Einstein's General Relativity.  In
contrast with the previous works, we consider the 
physically relevant case where the scalar field that breaks conformal symmetry 
and generates fermion masses is nonzero. In the physical gauge, in which this
scalar field is constant in space-time, the solution reproduces the weak-field
limit of the Schwarzschild--(anti)~De~Sitter solution modified by an 
additional term that, depending on the sign of the Weyl term in the action, is 
either oscillatory or exponential as a function of the radial distance.  Such 
behavior reflects the presence of, correspondingly, either a tachion or a 
massive ghost in the spectrum, which is a serious drawback of the theory under 
discussion.

\noindent PACS number(s): 04.50.+h
\end{abstract}

\bigskip

\noindent In a series of papers (see \cite{1,2,3,4,5,6,7,8} and
references therein), Mannheim and Kazanas explored the possibility that the
gravity is described by the conformally invariant theory with the key
ingredient in the action being the Weyl term
\beq \begin{array}{lcl}
I_{\rm W} &=& - \alpha \int d^4x \sqrt{-g}\, C_{\lambda\mu\nu\kappa}
C^{\lambda\mu\nu\kappa} \\  &=& - 2 \alpha \int d^4x \sqrt{-g}\, \left(
R_{\mu\nu} R^{\mu\nu} - \frac13 R^2 \right) + \mbox{boundary terms} ,
\end{array} \label{weyl}
\eeq
where $C^{\lambda}{}_{\mu\nu\kappa}$ is the conformal Weyl tensor and $\alpha$
is a purely dimensionless gravitational coupling constant (we use the system 
of units in which $\hbar = 1$ and $c = 1$). In particular, they obtained the
complete {\em conformally\/} static spherically symmetric solution \cite{1,6}
of the theory Eq.~(\ref{weyl}) with the line element given by
\beq
ds^2 = C^2(x)ds_0^2, \ \ \ ds_0^2 = - G(r)dt^2 + dr^2 / G(r) + r^2d\Omega,
\label{line}
\eeq
where $C(x)$ is an arbitrary nonzero function of the spacetime coordinates 
$x$, and $G(r)$ is given by
\beq
G(r) = 1 - \beta(2 - 3\beta\gamma)/r - 3\beta\gamma + \gamma r - \kappa r^2.
\label{solution}
\eeq
Here, $\beta$, $\gamma$, and $\kappa$ are integration constants.  
Having tacitly assumed that test bodies move along the
geodesics of the metric with the line element $ds_0^2$ of equation
(\ref{line}), Mannheim and Kazanas then claimed to recover the Newtonian term
($\propto 1/r$) in the potential of solution (\ref{solution}) of the conformal
gravity theory and also suggested \cite{1,5,7} that the additional
linear term $\gamma r$ in Eq.~(\ref{solution}) might account for the flat
galactic rotation curves without having to invoke dark matter.

However, solution (\ref{line}), (\ref{solution}) of the purely gravitational
conformal theory defined by Eq.~(\ref{weyl}) is not quite relevant to the
observations, since it is obtained without regard of the matter part of the
theory that includes the mass generation mechanism for the elementary 
particles and thereby for test bodies such as stars and planets. Such a 
feature of this solution is reflected in the unrestricted freedom of choosing 
the conformal factor $C(x)$ in Eq.~(\ref{line}) which clearly affects the 
timelike geodesics of the metric, but which is totally undetermined thus far.
Moreover, the electrovac generalization of solution (\ref{line}), 
(\ref{solution}) was previously obtained by Riegert \cite{Riegert}, who also 
asserted that one of the integration constants can be eliminated by further 
coordinate and conformal transformations.  This property of the solution, 
with $\gamma$ being such constant, was noted also in 
\cite{Schmidt1,Schimming} and very recently explicitly demonstrated in 
\cite{Schmidt2}.  All this makes very problematic the use of the metric given 
by the second expression in Eq.~(\ref{line}) and by Eq.~(\ref{solution}) as an 
observable one.  

 In the present paper, we consider this problem taking the matter to be 
represented by the following generic conformally invariant action \cite{3,4,6}
\beq
I_{\rm M} = - \int d^4x \sqrt{-g} \left[\partial^\mu S \partial_\mu S/2 +
\lambda S^4 - S^2 R / 12 + i \bar{\psi}\gamma^{\mu}(x) \nabla_\mu \psi - \zeta
S\bar{\psi}\psi\right], \label{matter}
\eeq
where $\psi$ is the fermion field, $S$ is the scalar field, $R$ is the
curvature scalar of the metric, and $\lambda$ and $\zeta$ are dimensionless
coupling constants. In the theory defined by Eqs.~(\ref{weyl}), 
(\ref{matter}), once the scalar field $S$ is everywhere nonzero it can be 
gauged to an identical constant $S_0$ by a conformal transformation. In this 
gauge, the fermion part of the action acquires the standard form with constant 
mass, hence all physical effects receive standard description; in particular,
massive particles and test bodies move along the timelike geodesics of the
metric as in General Relativity. It is clear that since conformal symmetry is
broken and there are massive particles in the real world, one should take
solutions with $S$ being nonzero. The physical vacuum is then regarded as the
state without excitations of the rest of the matter fields, in our case, the
field $\psi$.

 We consider solutions outside a compact source formed by the matter
fields (represented in our model by the single field $\psi$). The equations of
the theory, first introduced by Bach \cite{Bach} for the case of generic 
matter, have the form 
\beq
4 \alpha W_{\mu\nu} = T_{\mu\nu},  \label{eq}
\eeq
where the two sides stem, respectively, from the variation of the action
(\ref{weyl}) and the action (\ref{matter}) with respect to the metric, and the
expression of the stress-energy tensor $T_{\mu\nu}$ in the gauge $S \equiv 
S_0$ and with the $\psi$ field being zero is given by \cite{3,6}
\beq
T_{\mu\nu} = - S_0^2 \left(R_{\mu\nu} - g_{\mu\nu} R / 2\right)/6 - \lambda
S_0^4 g_{\mu\nu}. \label{t}
\eeq
Equations (\ref{eq}) with the right-hand side given by (\ref{t}) are nothing
but the Bach--Einstein equations with the cosmological constant term --- 
the last term in Eq.~(\ref{t}) (see, e.g., \cite{Schmidt3}). 
Note that the left-hand side of Eq.~(\ref{eq}) is identically traceless, and 
it is convenient to rewrite system (\ref{eq}) as
\beq
 4 \alpha W_{\mu\nu} = \T_{\mu\nu}, \ \ \  R = 24 \lambda S_0^2,
 \label{eqn}
\eeq
where $\T_{\mu\nu} \equiv - S_0^2 \left(R_{\mu\nu} - g_{\mu\nu} 
R/4\right)/6\,$ is the traceless part of the stress-energy tensor 
$T_{\mu\nu}$, and the second equation of system (\ref{eqn}) is the trace of 
equation (\ref{eq}).

We restrict ourselves to the static spherically symmetric case. As we 
explained above, we are interested in the situation where $T_{\mu\nu}$ is 
given by Eq.~(\ref{t}) with constant nonzero $S_0$. In this gauge, a static 
spherically symmetric metric can be put in the form
\beq
ds^2 = - B(r) dt^2 + A(r) dr^2 + r^2 d\Omega\, . \label{metric}
\eeq
Performing transformation of the radial coordinate as in \cite{1}, it is
convenient to rewrite this metric in the form
\beq
ds^2 = C^2(\rho) \left[ - D(\rho) dt^2 + d\rho^2 / D(\rho) + \rho^2 d\Omega
\right], \label{metricn}
\eeq
where $r^2 (\rho) = \rho^2 C^2(\rho)$, and the functions $C(\rho)$ and
$D(\rho)$ are simply related to $A(r)$ and $B(r)$ (see \cite{1}).

 Due to spherical symmetry and stationarity, the first set of equations
(\ref{eqn}) will give two independent equations which are conveniently chosen
to be
\beq
4 \alpha \left(W_0^0 - W_1^1\right) = \T_0^0 - \T_1^1\, , \ \ \ 4 \alpha 
W_{11} = \T_{11}\, , \label{eqnn}
\eeq
where index ``{\small 0}'' labels the time $t$ and index ``{\small 1}'' labels
the radial coordinate $\rho$. The expressions for the quantities in the
left-hand sides of equations (\ref{eqnn}) for the metric of 
Eq.~(\ref{metricn}) were obtained by Mannheim and Kazanas \cite{1,6}:
\beq
W_0^0 - W_1^1 = {D\left(\rho D\right)^{(4)} \over 3 \rho\, C^4}\, , 
\label{one}
\eeq \beq
W_{11} =  {1 \over 3\, C^2 D}\left({D^\prime D^{(3)} \over 2} -
{{D^{\prime\prime}}^2 \over 4} - {D D^{(3)} - D^\prime D^{\prime\prime} \over
\rho} - {D D^{\prime\prime} + {D^\prime}^2 \over \rho^2} + {2 D D^\prime \over
\rho^3} - {D^2 \over \rho^4} + \frac1{\rho^4}\right) , \label{two}
\eeq
where the superscript ``{\small ($n$)}'' indicates the $n$th derivative with
respect to the radial coordinate. Note that the function $C(\rho)$ has been
factored out in equations (\ref{one}) and (\ref{two}) due to the conformal
symmetry of the theory. The expressions for the right-hand sides of equations
(\ref{eqnn}) are
\beq
\T_0^0 - \T_1^1 = {S_0^2 D \over 3\, C} \left({C^\prime \over
C^2}\right)^\prime \, ,
\eeq \beq
\T_{11} = - {S_0^2 \over 12} \left({D^{\prime\prime} \over 2 D} + {1 - D \over
\rho^2 D} + 3 F^{\prime\prime} + {D^\prime F^\prime \over D} - {2 F^\prime
\over \rho} - 3 {F^\prime}^2 \right)\, ,
\eeq
where $F = \log C$. The scalar curvature is given by
\beq
R = {6 (\rho^2 D C^\prime)^\prime - C \left(\rho^2 (1 - D)
\right)^{\prime\prime} \over \rho^2 C^3} \, .
\eeq

In a general case, it appears to be difficult to obtain an exact solution for
$C(\rho)$ and $D(\rho)$. However, it is possible to obtain the weak-field 
limit of the solution. Let the physical metric of Eq.~(\ref{metric}) in the 
spatial region of interest be sufficiently close to the flat one, so that
\beq
A(r) = 1 + \epsilon a(r), \ \ \ B(r) = 1 - \epsilon b(r), \label{ab}
\eeq
where $\epsilon$ is an auxiliary small parameter to be set equal to unity in
the end. Then, in the same region, the functions $C(\rho)$ and $D(\rho)$ of 
the metric of Eq.~(\ref{metricn}) and the radial coordinate transformation
$r(\rho)$ to the first order in $\epsilon$ are given by
\beq
C(\rho) = 1 + \epsilon f (\rho), \ \ \ D(\rho) = 1 - \epsilon h (\rho), \ \ \
r(\rho) = \rho [1 + \epsilon f(\rho)], \label{cd}
\eeq
with the functional relation
\beq
a(r) = h(r) - 2 r f^\prime (r), \ \ \ b(r) = h(r) - 2 f (r). \label{relation}
\eeq

To obtain the system of equations for the functions $h(\rho)$ and $f(\rho)$, 
we linearize the equations of system (\ref{eqn}) for the metric of
Eqs.~(\ref{metricn}),~(\ref{cd}) in the small parameter $\epsilon$. First, we
note that the scalar curvature $R$ of this metric is of order $\epsilon$.
Hence, the second equation of system (\ref{eqn}) implies that the 
dimensionless value of $\lambda S_0^2 r^2$ should also be at least of order 
$\epsilon$ in the spatial region under consideration. On the observational 
grounds, this restriction on the value of $\lambda S_0^2 r^2$ is quite natural 
since this value represents the effect of the cosmological constant which is 
believed to be small on the galactic and stellar spatial scales. However, from 
the theoretical viewpoint, such a restriction constitutes the fine-tuning 
problem of the cosmological constant. The solution of this long-standing 
problem is absent, so we formally replace $\lambda$ by $\epsilon \lambda$, 
thus taking into account the smallness of the corresponding parameter.

It remains to linearize equations (\ref{eqnn}) and the last equation of system
(\ref{eqn}) in the small parameter $\epsilon$. Omitting the simple but
cumbersome calculations, we present the corresponding result in the form of 
the following system for the functions $h(\rho)$ and $f(\rho)$:
\beq
- \left(\rho h \right)^{(4)} = 6 p \rho f^{\prime\prime} , \label{onen}
\eeq \beq
{1 \over 3\rho}\left(h^{(3)} + {h^{\prime\prime} \over \rho} - {2 h^\prime
\over \rho^2} + {2 h \over \rho^3} \right) = \frac{p}{2}\left(
{h^{\prime\prime} \over 2} - \frac{h}{\rho^2} - 3 f^{\prime\prime} + {2
f^\prime \over \rho} \right)\, , \label{twon}
\eeq \beq
6 \left(\rho^2 f^\prime\right)^\prime = \left(\rho^2 h\right)^{\prime\prime} +
24 q \rho^2\, , \label{threen}
\eeq
where we made the notation
\beq
p = {S_0^2 \over 24 \alpha}\, , \ \ \ q = \lambda S_0^2\, . \label{qp}
\eeq

We proceed to the solution of these equations. First, it is convenient to set
\beq
h(\rho) = 2 m /\rho - 2 q \rho^2 + v(\rho)\, , \label{v}
\eeq
where $v(\rho)$ is the new unknown function and $m$ is a constant. Equations
(\ref{onen}) and (\ref{threen}) can be integrated once. The integration
constant that appears after the integration of Eq.~(\ref{onen}) can be set to
any value by rescaling the time and length which shifts $f(\rho)$ by a
constant. We use this property to eliminate the parameter $q$ from the
equations for $v(\rho)$ and $f(\rho)$. The integration constant that appears
after the integration of Eq.~(\ref{threen}) can be eliminated by the
redefinition of the constant $m$ in Eq.~(\ref{v}). As a result, we obtain the
following system of equations:
\beq
\rho v^{(3)} + 3 v^{\prime\prime} + 6 p \left(\rho f^\prime - f \right) = 
0\, , \label{onenn}
\eeq \beq
v^{(3)} + {v^{\prime\prime} \over \rho} - \frac{3p}4 \, \rho v^{\prime\prime} 
- {2 v^\prime \over \rho^2} + {\left(4 + 3 p \rho^2\right) \over 2 \rho^3} v +
\frac{9p}2\, \rho f^{\prime\prime} - 3p f^\prime = 0 \, , \label{twonn}
\eeq \beq
v^\prime + {2 v / \rho} - 6 f^\prime = 0 \, . \label{threenn}
\eeq

Now it is convenient to proceed to the new unknown functions $\tilde a(\rho)$
and $\tilde b(\rho)$ related to $v(\rho)$ and $f(\rho)$ as follows 
[cf. Eq.~(\ref{relation})]:
\beq
\tilde a(\rho) = v(\rho) - 2 \rho f^\prime (\rho)\, , \ \ \  \tilde b(\rho) =
v(\rho) - 2 f (\rho) \, . \label{tilde}
\eeq
From system (\ref{onenn})--(\ref{threenn}), one easily obtains the following
system of equations for $\tilde a(\rho)$ and $\tilde b(\rho)$:
\beq
\tilde b^{\prime\prime} + {2 \tilde b^\prime / \rho} + p\, \tilde b = 0\, , 
\ \ \ \rho\, \tilde b^\prime + 2 \tilde a = 0\, .
\eeq
Its solution is straightforward and depends on the sign of the constant $p$
that coincides with the sign of $\alpha$ [see Eq.~(\ref{qp})].  First, we
consider the case where  $p > 0$.  We obtain
\beq
\tilde a(\rho) = n \left[{\sin\left( k \rho + \phi\right) \over \rho} - k \cos
\left( k \rho + \phi\right) \right] \, , \label{tildea}
\eeq \beq
\tilde b(\rho) = {2 n \sin\left( k \rho + \phi\right) \over \rho} \, ,
\label{tildeb}
\eeq
where $k = \sqrt p$, and $n$ and $\phi$ are integration constants. Combining
Eqs.~(\ref{relation}), (\ref{v}), (\ref{tilde}), and (\ref{tildea}),
(\ref{tildeb}), we eventually obtain the following solution of our problem:
\beq
a(r) = {2 m \over r} - 2 q r^2 + n \left[{\sin\left( k r + \phi\right) 
\over r} - k \cos \left(k r + \phi \right) \right] \, , \label{fina}
\eeq \beq
b(r) = {2 m + 2 n \sin\left(k r + \phi\right) \over r} - 2 q r^2  \, .
\label{finb}
\eeq
We see that in the Newtonian limit, apart from the universal term $q r^2$,
there arises the additional gravitational potential
\beq
V(r) = {}- {m + n \sin\left( k r + \phi\right) \over r}\, , \label{pot}
\eeq
in which the constants $m$, $n$, and $\phi$ are to be related to the source.
The constants $k = \sqrt p$ and $q$ are universal and are given by
Eq.~(\ref{qp}).  

 We note that the linearized static spherically symmetric solutions in a 
generic (not conformally invariant) second-order gravitational 
theory without the cosmological constant were obtained 
in \cite{Stelle}. Their structure is similar to that of (\ref{fina}), 
(\ref{finb}) and to the solutions (\ref{expona}), (\ref{exponb}) 
below.  However, it is not possible to pass to a direct 
limit of conformal invariance in the solutions of \cite{Stelle},
because the case of conformal invariance is characterized by a nontrivial 
degeneracy, in particular, the massive scalar degree of freedom which is 
present in a generic case is missing here (see also \cite{Schmidt3} in 
this respect).

Now suppose that a static compact source is composed of identical ``atoms''
(these may be real atoms or elementary particles) and that each of these atoms
produces static gravitational potential as given by Eq.~(\ref{pot}) with
identical constants $m$, $n$, and $\phi$. In view of the weakness of the
potential, we also assume the validity of the superposition principle. Then, 
if $\mu (\r)$ is the spatial distribution of the ``atoms'' in the source, the
total potential is given by the expression
\beq
\Phi (\r) = \int V\left(\left|\r - \r^\prime\right|\right) \mu
\left(\r^\prime\right) d^3 \r^\prime . \label{Phi}
\eeq
This potential is the sum of two terms: $\Phi (\r) = \Phi_m (\r) + \Phi_n
(\r)$.  They satisfy the equations
\beq\begin{array}{l}
\Delta \Phi_m (\r) = 4 \pi m \mu (\r), \\ \Delta \Phi_n (\r) + p\, \Phi_n (\r)
= 4 \pi n \sin \phi\, \mu(\r),
\end{array} \label{Poisson} \eeq
that, in the theory under investigation, correspond to the unique Poisson
equation of the linearized General Relativity.

For a spherically symmetric compact distribution $\mu (r)$, the potential 
given by Eq.~(\ref{Phi}) with the kernel given by Eq.~(\ref{pot}) is easily
calculated:
\beq
\Phi (r) = - \int_r^\infty {M(r^\prime) \over {r^\prime}^2} d r^\prime  - {N
\sin( k r + \phi) \over r} - {4 \pi n \sin \phi \over k r} \int_r^\infty \mu
(r^\prime) \sin \left[k \left(r - r^\prime \right) \right] r^\prime d r^\prime
, \label{Phi1}
\eeq
where
\beq
M(r) = 4 \pi m \int_0^r \mu (r^\prime) {r^\prime}^2 d r^\prime, \ \ \ N = {4
\pi n \over k} \int_0^\infty \mu (r^\prime) \sin \left(k r^\prime \right)
r^\prime d r^\prime .  \label{MN}
\eeq
Thus, outside the source, the potential of the form (\ref{pot}) is reproduced
with the same phase $\phi$, but with different coefficients $m$ and $n$.
Moreover, while the coefficient $m$ is additive (it plays the role of the
gravitational mass of the source), the coefficient $n$ is not: its new value
$N$ is given by the second expression in Eq.~(\ref{MN}). However, the
coefficient $n$ becomes approximately additive for a distribution whose 
spatial size is significantly less than $1/k$.

If the product $k r < 1$ in the region of interest (say, on galactic scales),
one can expand the oscillatory part of Eq.~(\ref{pot}) in powers of $k r$ to
obtain
\beq
V(r) = V_0 - {M_0 \over r} + {\Gamma r \over 2} + Q r^2 + {\cal O} \left[
\left( k r \right)^3\right] \, , \label{expand}
\eeq
where $V_0 = - n k \cos \phi\,$, $M_0 = m + n \sin \phi\,$, $\Gamma = n k^2
\sin \phi\,$, and $Q = q + n k^3 \cos \phi / 6\,$. We thus recover the linear
term in the potential of Eq.~(\ref{expand}), similar to that which was used by
Mannheim and Kazanas \cite{1,5,7} to account for the flat galactic 
rotation curves.  However, there exists an important observational bound that 
rules out the possibility for the linear term in the expansion (\ref{expand}) 
to play a significant role on galactic scales. Note that the coefficients 
$-g_{00} (r)$ and $g_{rr} (r)$ of the metric of our solution are not mutually 
inverse, which is reflected in the fact that the functions $a(r)$ and $b(r)$, 
given, respectively, by Eqs.~(\ref{fina}) and (\ref{finb}), are not equal to 
each other.  At small enough distances, both functions reproduce Newtonian
potentials with the masses, respectively, $m_0 = m + n \sin \phi$ and $m_1 = m
+ (n \sin \phi) / 2$, the difference between them being $\Delta m = (n \sin
\phi) / 2$. At the same time, the {\em Viking\/} spacecraft observations in 
the vicinity of the Sun indicate that the ratio $\Delta m / m \lsim 2 \times
10^{-3}$ \cite{Reasenberg} (see also \cite{Will}). This implies the following
observational bound for the Sun:
\beq
{n \sin \phi \over m} \lsim 4 \times 10^{-3} . \label{bound}
\eeq
Since we assume that the parameter $k$ is sufficiently small so that the
expansion (\ref{expand}) is legitimate on galactic scales, the values of both
$m$ and $n$ are additive on such scales and the estimate (\ref{bound}) 
is valid on galactic scales as well.  Now, the linear term in
Eq.~(\ref{expand}) formally becomes comparable in magnitude to the Newtonian
one only at the distance $r \sim \sqrt{M/\Gamma} \approx \sqrt{m / (n k^2 \sin
\phi)}$. But, for such distances, we would have $k r \sim \sqrt{m / (n \sin
\phi)} \gsim 10$ because of estimate (\ref{bound}), which contradicts the
original assumption $k r < 1$. Thus, the linear term in the expansion
(\ref{expand}) cannot play a significant role on galactic scales, and one
should rather try the whole potential in the form (\ref{Phi1}) for a
spherically symmetric source with the bound (\ref{bound}) to account for the
galactic rotation curves.  Such a possibility still remains to be 
investigated.

It is instructive to estimate the realistic value of the constant $\alpha$ in
Eq.~(\ref{weyl}) for which the value of $k r$ is of order unity on a typical
galactic scale of $10$~kpc, thus making the potential of the form (\ref{pot})
in principle relevant to the galactic rotation curves.  Whatever scalar fields
are present in the theory, they all contribute to the value of $p$ given
by Eq.~(\ref{qp}). Thus, at least the scalar Higgs field of the Standard Model
of strong and electroweak interactions should be taken into account. The mean
value of this field is known to be $\eta \simeq 246$~Gev, this value will
contribute to $S_0$ in Eq.~(\ref{qp}) and, in order that $k \, \times (10~{\rm
kpc}) \lsim 1$ be valid, we must have
\beq
\alpha \gsim 10^{74} \, , \label{estim}
\eeq
which, of course, is a severe restriction. It is possible to conceive models 
in which this restriction is weakened; for instance, one can introduce
another scalar field with the ``wrong'' overall sign in the action, so that
its contribution to the value of the parameter $p$ will be of the opposite 
sign.  Its vacuum expectation vaue has to be fine-tuned to counterbalance the
contribution of $\eta$. Another possibility is to try to construct particle
theory without fundamental scalar fields, although it may turn out to be very
difficult to do this in a {conformally invariant\/} manner.

On the other hand, if we take $\alpha \sim 1$, then the expectation value 
$\eta \simeq 246$~Gev of the Standard Model Higgs field leads to the spatial
scale
\beq
1 / k \sim 10^{- 16}~{\rm cm}
\eeq
on which the potential (\ref{pot}) oscillates.  Its significance might only be
manifest on the spatial scales of elementary particles, where, of course, the
whole theory must be quantized.

In the case of $p < 0$, which corresponds to $\alpha < 0$, the solution for 
$a (r)$ and $b (r)$ has the form  
\beq
a (r) = {2 m \over r} - 2 q r^2 + n_1 \left(1 + k r \right) {e^{- k r} 
\over r} + n_2 \left(1 - k r \right) {e^{k r} \over r} , \label{expona}
\eeq \beq
b (r) = {2 m \over r} - 2 q r^2 +  2 n_1 {e^{- k r} \over r} + 2 n_2 {e^{k r}
\over r} , \label{exponb}
\eeq
where now $k = \sqrt{-p}\,$, and $n_1$ and $n_2$ are integration constants. 
Similar solutions in a generic second-order gravitational theory 
(not conformally invariant) without the cosmological constant 
were obtained in \cite{Stelle}. Solutions in the conformally invariant 
second-order theory with the Einstein term but without the 
cosmological-constant term 
were also obtained in \cite{Schmidt3}.  The physically meaningful 
solution is selected by imposing boundary conditions at infinity, what 
leads to the condition $n_2 = 0$. For sufficiently small values of $k$, the 
observational bound similar to (\ref{bound}) implies
\beq
{n_1 \over m} \lsim 4 \times 10^{-3}, \label{bound1}
\eeq
and makes the extra exponential potential in (\ref{exponb}) uninteresting.

Finally, we note that in the case of $p < 0$, which corresponds to $\alpha
< 0$, one can also obtain solutions by formally replacing the trigonometric 
functions in Eqs.~(\ref{fina}), (\ref{finb}), and (\ref{pot}) by their 
hyperbolic counterparts and taking $k = \sqrt{-p}$.  Equations (\ref{Poisson}) 
will then remain valid in this case as well, with the replacement of 
$\sin \phi$ by $\sinh \phi$. The structure of the left-hand sides of equations 
(\ref{Poisson}) reflects, besides the presence of the massless graviton, also 
the well-known presence of a spin-two tachion (in the case of $\alpha > 0$) or 
a spin-two massive ghost (in the case of $\alpha < 0$) on the background with 
$S \ne 0$ of the theory described by Eqs.~(\ref{weyl}),~(\ref{matter}) (see
\cite{Stelle,JT,Riegert} and references therein). The presence of a tachion 
in the case of $\alpha > 0$ indicates instability of a large class of classical
solutions, including the flat space-time solution in the case of 
$\lambda = 0$; and the presence of a ghost in the case of $\alpha < 0$ implies 
possible absence of perturbative unitarity in the corresponding quantum 
theory. This appears to be the main drawback of the conformal theory under 
discussion.

The authors are grateful to Professor~P.~Mannheim for valuable discussions 
and to the referee for drawing their attention to several papers.   
This work was supported in part by the Foundation of Fundamental Research of
the Ministry of Science of Ukraine under grant No.~2.5.1/003.


\begin{thebibliography}{99}

\bibitem{1}
P.~D.~Mannheim and D.~Kazanas, Ap.\ J., {\bf 342}, 635 (1989).

\bibitem{2}
P. D. Mannheim, {\it Gen. Rel. Grav.} {\bf  22}, 289 (1990);
D.~Kazanas and P.~D.~Mannheim, Ap.\ J.\ Suppl.\ Ser.\ {\bf 76}, 431 (1991);
P.~D.~Mannheim and D.~Kazanas, Phys.\ Rev.\ D {\bf 44}, 417
(1991).

\bibitem{3}
P.~D.~Mannheim, Ap.\ J.\  {\bf 391}, 429 (1992).

\bibitem{4}
P.~D.~Mannheim, Gen.\ Rel.\ Grav.\ {\bf 25}, 697 (1993).

\bibitem{5}
P.~D.~Mannheim, Ap.\ J.\  {\bf 419}, 150 (1993).

\bibitem{6}
P.~D.~Mannheim and D.~Kazanas, Gen.\ Rel.\ Grav., {\bf 26}, 337 (1994).

\bibitem{7}
P.~D.~Mannheim, Found.\ Phys.\ {\bf 26}, 1683 (1996);
P.~D.~Mannheim, Ap.\ J. {\bf 479}, 659 (1997).

\bibitem{8}
P.~D.~Mannheim, Phys.\ Rev.\ D {\bf 58}, 103511 (1998).

\bibitem{Riegert}
R.~J.~Riegert, Phys.\ Rev.\ Lett.\ {\bf 53}, 315 (1984).

\bibitem{Schmidt1}
H.-J.~Schmidt, Ann.\ Phys.\ (Leipz.) {\bf 41}, 435 (1984).

\bibitem{Schimming}
R.~Schimming, ``On the Bach and the Bach--Einstein gravitational field 
equations,'' in: Proc.\ of the Int.\ Seminar ``Current Topics in 
Mathematical Cosmology,'' Potsdam, Germany, 30~March--4~April, 1998, 
ed.\ by M.~Rainer and H-J.~ Schmidt (World Scientific, Singapore, 1998), 
pp.~39--46.  

\bibitem{Schmidt2}
H.-J.~Schmidt, Preprint UNIPO-MATH-99-May-27, gr-qc/9905103.

\bibitem{Bach}
R.~Bach, Math.\ Zeitschr.\ {\bf 9}, 110 (1921).

\bibitem{Schmidt3}
H.-J.~Schmidt, Astron.\ Nachr.\ {\bf 306}, 67 (1985);  
H.-J.~Schmidt, {\it ibid}. {\bf 306}, 231 (1985); 
H.-J.~Schmidt, {\it ibid}. {\bf 307}, 339 (1986). 

\bibitem{Stelle}
K.~S.~Stelle, Gen.\ Rel.\ and Grav.\ {\bf 9}, 353 (1978). 
There appears to be an error in Eq.~(3.6b) of this reference: the terms 
$\propto r^{-1} \exp\left(\pm m_2 r\right)$ should be multiplied by $1/2$; 
see the second reference in \cite{Schmidt3}. 

\bibitem{Reasenberg}
R.~D.~Reasenberg {\em et al}., Ap.\ J.\ Lett.\ {\bf 234}, 219 (1979).

\bibitem{Will}
C.~M.~Will, {\em Theory and Experiment in Gravitational Physics\/} (Cambridge
University Press, Cambridge, 1993).

\bibitem{JT}
J.~Julve and M.~Tonin, Nuovo Cim.\ {\bf 46B}, 137 (1978).
 
\end{thebibliography}
\end{document}